# UADSN: Uncertainty-Aware Dual-Stream Network for Facial Nerve Segmentation

Guanghao Zhu
*School of Optoelectronic Science & Engineering*
*University of Electronic Science and Technology of China*
Chengdu, China
202222050412@std.uestc.edu.cn

Lin Liu
*School of Optoelectronic Science & Engineering*
*University of Electronic Science and Technology of China*
Chengdu, China
liulin1979@uestc.edu.cn

Jing Zhang
*School of Optoelectronic Science & Engineering*
*University of Electronic Science and Technology of China*
Chengdu, China
zhangjing@uestc.edu.cn

Xiaohui Du
*School of Optoelectronic Science & Engineering*
*University of Electronic Science and Technology of China*
Chengdu, China
xiaohuie@126.com

Ruqian Hao
*School of Optoelectronic Science & Engineering*
*University of Electronic Science and Technology of China*
Chengdu, China
ruqian_hao@163.com

Juanxiu Liu*
*School of Optoelectronic Science & Engineering*
*University of Electronic Science and Technology of China*
Chengdu, China
juanxiul@uestc.edu.cn

*Abstract*—Facial nerve segmentation is crucial for preoperative path planning in cochlear implantation surgery. Recently, researchers have proposed some segmentation methods, such as atlas-based and deep learning-based methods. However, since the facial nerve is a tubular organ with a diameter of only 1.0-1.5mm, it is challenging to locate and segment the facial nerve in CT scans. In this work, we propose an uncertainty-aware dual-stream network (UADSN). UADSN consists of a 2D segmentation stream and a 3D segmentation stream. Predictions from two streams are used to identify uncertain regions, and a consistency loss is employed to supervise the segmentation of these regions. In addition, we introduce channel squeeze & spatial excitation modules into the skip connections of U-shaped networks to extract meaningful spatial information. In order to consider topology-preservation, a clDice loss is introduced into the supervised loss function. Experimental results on the facial nerve dataset demonstrate the effectiveness of UADSN and our submodules.

## I. Introduction

Cochlear implantation surgery is a surgical procedure performed to treat hearing loss, which requires drilling from the skull surface to the cochlea during surgery [1]. To minimize invasion, it is essential to identify critical structures around the cochlea for preoperative path planning. Among these structures, the facial nerve is of particular importance. It is a tubular structure with a diameter of approximately 1.0-1.5mm, controlling all movements of the ipsilateral facial muscles. Damage to the facial nerve during surgery may lead to temporary or permanent facial paralysis. Temporal bone Computed Tomography (CT) data provide surgeons with anatomical information of the patient for determining the orientation and geometry of the facial nerve during preoperative planning. However, due to the small volume and blurry boundaries of the facial nerve, manual segmentation is a time-consuming and labor-intensive task, requiring sophisticated knowledge of radiology and surgical anatomy.

In facial nerve segmentation, traditional methods often rely on atlas-based approaches [1]-[5]. Firstly, the prior shape information of the target tissue is obtained to form an atlas, and then the established atlas is registered with the image to be segmented. The atlas with a high matching degree is transformed to generate the segmentation results. However, the segmentation effect of the atlas-based method heavily depends on the quality of the established atlas and registration process. When there are significant anatomical variations, registration errors are substantial, which may lead to serious segmentation errors.

In recent years, with the rapid development of deep learning, medical image segmentation research has made great breakthroughs. Many segmentation networks have been proposed, including CNN-based [6]-[9] and Transformer-based methods [10]. For 2D medical image segmentation, U-Net [6], which uses skip connections between the encoder and decoder, has become the main architecture. For 3D medical image segmentation, 2D segmentation methods employ a slicing strategy to segment images, making it difficult to fully utilize the 3D spatial information of medical volumetric images, resulting in discontinuous segmentation. Therefore, some 3D networks have been proposed, such as 3D U-Net [7], which replaces the 2D operations in 2D U-Net with their 3D counterparts, significantly improving the accuracy of anatomical structure segmentation in medical volumetric data. For facial nerve segmentation, Nikan et al. [11] proposed a patch-wise densely connected 3D network called PWD-3DNet, achieving a Dice score of 0.74 for facial nerve. However, 3D networks typically have more parameters and computational costs, making them prone to overfitting on small-scale datasets.

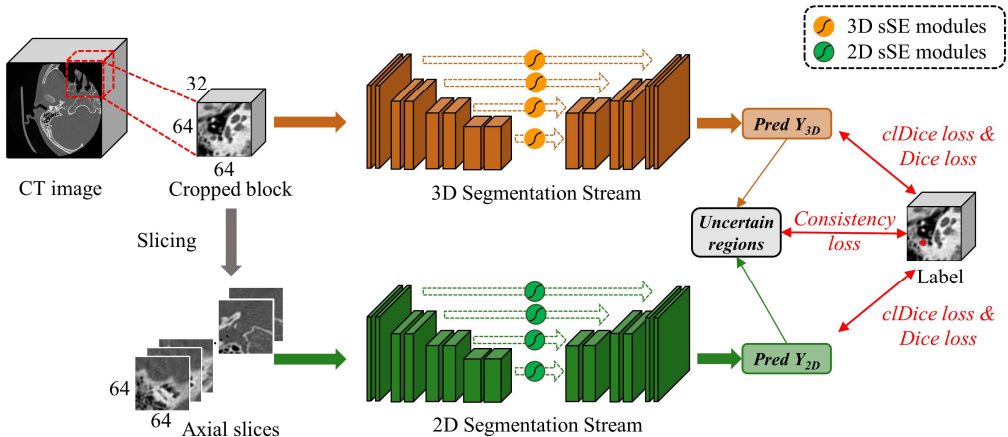

Fig. 1. The framework of our proposed UADSN. The UADSN consists of two segmentation streams: a 3D segmentation stream and a 2D segmentation stream.

In this paper, we propose an uncertainty-aware dual-stream network (UADSN) for facial nerve segmentation. The UADSN consists of a 2D segmentation stream and a 3D segmentation stream. The input to the 3D segmentation stream consists of sub-volumes randomly sampled from temporal bone CT scans, while the input to the 2D segmentation stream consists of the corresponding axial CT slices of these sub-volumes. Regions where the predictions of the two streams differ are considered to be uncertain areas. Under the supervision of consistency loss over uncertain regions, the network's accuracy improves in the region of the edge of the facial nerve. Given that the facial nerve is a tiny tubular structure, so it is important and challenging to maintain the connectivity of the topology of the segmentation result. However, traditional volume-based metrics, such as Dice, primarily measure the overlap between two regions and are suboptimal for topology-preservation. Therefore, we introduce the centerlineDice (clDice) loss [12] to ensure better connectivity of the segmentation. Additionally, the channel squeeze & spatial excitation (sSE) modules [13] are employed in skip connections to highlight significant spatial information. In summary, the contributions of this paper can be summarized as follows:

(1) We propose UADSN, a network that combines 2D and 3D CNNs for facial nerve segmentation. UADSN leverages the predictions from both 2D and 3D CNNs to automatically identify uncertain regions and uses consistency loss to enhance the segmentation accuracy in edge areas.

(2) The clDice loss [12] is used to maintain the topological connectivity of the facial nerve segmentation results. The sSE modules [13] are used to introduce the spatial attention mechanism.

(3) We conduct extensive experiments on our self-built facial nerve dataset. The experimental results indicate the superior performance of UADSN compared to other popular segmentation methods. Additionally, ablation studies demonstrate the effectiveness of each component.

## II. RELATED WORK

### A. Atlas-Based Segmentation Methods

In the atlas-based method, prior anatomical information is extracted first, and then a registration process is performed to estimate the anatomical differences between the atlas and the target image. If the atlas closely matches the subject's anatomy, segmentation is achieved by matching the transformed atlas labels to the objects [14]. For the segmentation of the facial nerve, Reda et al. [2] extracted the centerline using a minimal cost path algorithm and expanded it into the full structure with a level-set algorithm. Gare et al. [5] utilized high-resolution micro-CT scans to create atlases of the facial nerve. To approximate the centerline of the facial nerve, the user is required to place four fiducials on the CT image. The atlas that is the closest match to the centerline is selected for registration and produces the segmentation. However, identifying the four fiducials demands extensive training and knowledge of radiological.

### B. Deep Learning-Based Segmentation Methods

In recent years, medical image segmentation methods based on deep learning have developed rapidly. U-Net architecture has been widely used, and researchers have proposed various modifications and extensions. For instance, Zhou et al. [9] proposed U-Net++, which redesigns the skip pathways and introduces deep supervision. Ali et al. [10] introduced UNETR, utilizing a transformer as the encoder to capture the global multi-scale information. As for the segmentation of facial nerve, Lv et al. [15] proposed a lightweight 3D CNN named W-Net. They applied the maximum region growing approach to eliminate segmentation errors and achieved a Dice score of 0.77. Dong [16] et al. proposed a 2D network called FNSegNet, which can be divided into two stages, i.e. coarse and fine segmentation stages. In the coarse segmentation stage, the rough region of the facial nerve is predicted by expanding the receptive field. In the fine segmentation stage, the features are separated into center and boundary features and optimized using a spatial attention module (SAM) to refine the segmentation. Ding et al. [17] employed no new U-Net (nnU-Net) [18] to segment facial nerve, which can automatically self-configure to new datasets.

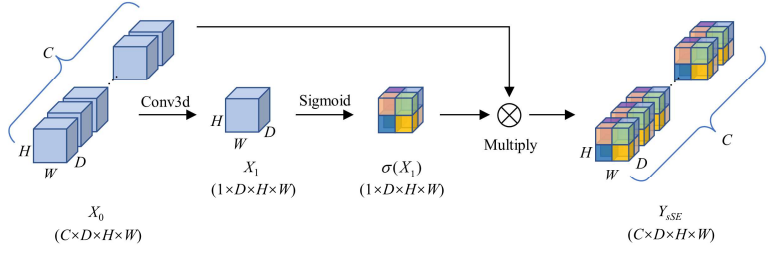

Fig. 2. 3D channel squeeze & spatial excitation (sSE) module.

However, existing works have not focused on the topological connectivity of facial nerve segmentation results, mostly using volume-based loss functions such as Dice loss. Therefore, in our work, the clDice loss [12] is used to guarantee topology-preservation, which is crucial for the segmentation of tiny tubular organs like the facial nerve.

## III. METHODS

The architecture of UADSN is shown in Fig. 1. It can be seen that UADSN consists of a 2D segmentation stream and a 3D segmentation stream. Their inputs are the sub-volumes sampled from temporal bone CT scans and the corresponding axial slices, respectively.

### A. Uncertainty-Aware Dual-Stream Structure

Since the ratio of facial nerve region is very small in CT images, it is difficult to perform facial nerve segmentation directly in complete CT slices with a 2D CNN. Therefore, we crop blocks of size $64\times64\times32$ along the X-axis, Y-axis, and Z-axis, and then input the blocks into a 3D CNN. We use a 3D U-Net structure as the segmentation sub-network, and the segmentation result is defined as $Y_{3D}$.

Although 2D CNN is unable to explore the inter-slice connection, it has fewer parameters compared to 3D CNN and can provide a larger receptive field. Therefore, we use a 2D U-Net architecture for the 2D segmentation stream, with axial slices of the input blocks as input. The outputs are concatenated to get the result of facial nerve segmentation in the block, denoted as $Y_{2D}$.

To achieve fine-grained segmentation in regions that are difficult to segment, such as boundary regions, we identify uncertain regions based on the results of the two segmentation streams. Specifically, we first perform binary processing on $Y_{3D}$ and $Y_{2D}$ to obtain $\tilde{Y}_{3D}$ and $\tilde{Y}_{2D}$. Using the XOR operation, the mask of the uncertain region $M$ is obtained as follows:

$$M = \tilde{Y}_{3D} \oplus \tilde{Y}_{2D}, \quad (1)$$

where $\oplus$ denotes the XOR operation. Then, we design uncertainty-aware consistency loss $L_c$ as the voxel-level mean squared error (MSE) loss between the prediction result and the label on the uncertain region:

$$L_c = \left[ \frac{\sum_i M_i * \left\| Y_{2D}^i - Y^i \right\|^2}{\sum_i M_i} + \frac{\sum_i M_i * \left\| Y_{3D}^i - Y^i \right\|^2}{\sum_i M_i} \right] / 2, \quad (2)$$

where $i$ denotes the $i-th$ voxel; $Y^i$ denotes the value of the label at the $i-th$ voxel; $Y_{2D}^i$ and $Y_{3D}^i$ denote the prediction of the 2D and 3D segmentation streams at the $i-th$ voxel. With our uncertainty-aware consistency loss, the network gradually focuses on difficult regions during training to achieve fine-grained segmentation.

### B. Spatial Attention Modules

In U-Net, skip connections are used to fuse high-level features with low-level features. In order to extract meaningful spatial information from low-level features while suppressing unimportant information, we employ 2D and 3D channel squeeze & spatial excitation (sSE) modules [13] in the skip connections of 2D segmentation stream and 3D segmentation stream, respectively. A 3D version of the sSE module is shown in Fig. 2. We consider the input feature map $X_0 \in \mathbb{R}^{C \times D \times H \times W}$, where $C, D, H, W$ denote channel, depth, height, and width dimensions. A 3D convolution operation is applied to squeeze the channel dimension, generating a projection tensor $X_1 \in \mathbb{R}^{1 \times D \times H \times W}$. Then a sigmoid activation operation is used to rescale tensor to $[0,1]$. $X_0$ is excited across the spatial dimension by an element-wise multiplication, which is defined as follows:

$$Y_{sSE} = \sigma(X_1) \otimes X_0, \quad (3)$$

where $\otimes$ denotes element-wise multiplication. The sSE module helps the network focus more on relevant spatial information and ignore the irrelevant ones.

### C. ClDice Loss

To ensure the topological connectivity of the facial nerve segmentation, we use clDice [12] as a connectivity-preserving metric to evaluate the segmentation of the tubular facial nerve. First, the 3D parallel thinning algorithm [19] is used to extract the facial nerve skeletons $S_L$ and $S_P$ from ground truth masks ($V_L$) and predicted masks ($V_P$). Subsequently, we compute the Topology Precision Tprec and Topology Sensitivity Tsens, defined as follows:

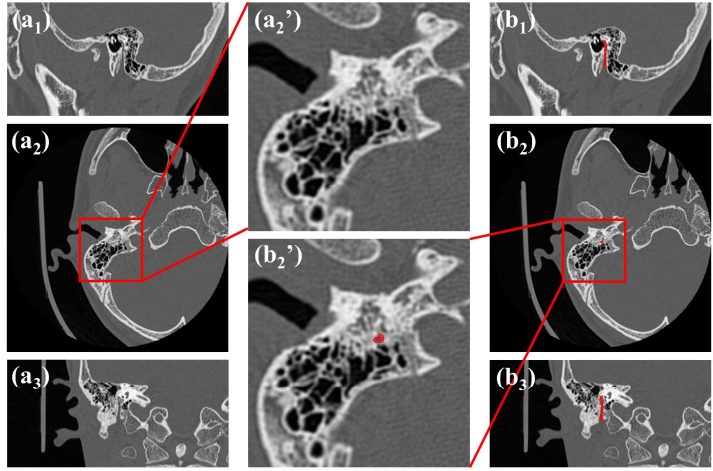

Fig. 3. Example CT slices in our dataset. The region where the facial nerve is located is marked in red.

$$\mathrm{Tprec}(S_P, V_L) = \frac{|S_P \cap V_L|}{|S_P|}, \quad (4)$$

$$\mathrm{Tsens}(S_L, V_P) = \frac{|S_L \cap V_P|}{|S_L|}. \quad (5)$$

Since we aim to maximize both precision and sensitivity, clDice is defined as the harmonic mean of the two metrics:

$$clDice(V_P, V_L) = 2 \times \frac{\mathrm{Tprec}(S_P, V_L) \times \mathrm{Tsens}(S_L, V_P)}{\mathrm{Tprec}(S_P, V_L) + \mathrm{Tsens}(S_L, V_P)}. \quad (6)$$

To achieve accurate facial nerve segmentation while preserving the tubular structure, we combine Dice loss and clDice loss. Thus, the total loss function of the proposed UADSN is defined as:

$$L = (1-\alpha)(1-Dice) + \alpha(1-clDice) + \beta L_c, \quad (7)$$

where weight $\alpha$ is used to trade-off volume-based performance and topology-preservation; $\beta$ is a time-dependent Gaussian warming up function [20] defined as follows:

$$\beta(t) = 0.1 * \exp(-5(1 - t/t_{max})^2), \quad (8)$$

where $t$ denotes the current iteration number and $t_{max}$ represents the maximum number of iterations. $\beta$ gradually increases during training, so the loss function is initially dominated by the Dice loss and the clDice loss, allowing the network to perform meaningful facial nerve segmentation. In the later stages of training, the network will focus on difficult-to-segment regions to achieve fine-grained segmentation.

## IV. EXPERIMENTS

### A. Datasets and Pre-Processing

Since there is no publicly available facial nerve dataset, we use our self-built facial nerve dataset to test the segmentation performance of UADSN. The dataset contains 28 temporal bone CT images collected at the First Affiliated Hospital of Guangzhou Medical University. The facial nerves have been voxel-wise annotated by clinically experienced doctors. The size of each temporal bone CT image ranges from $512 \times 512 \times 90$ to $512 \times 512 \times 146$. The voxel size of the CT image is $0.293 \times 0.293 \times 0.625$ mm$^3$. We split the 28 cases into 23 for training and 5 for evaluation. Fig. 3 shows example CT slices. Fig. 3($a_1$), ($a_2$), and ($a_3$) represent the sagittal, axial, and coronal view of the temporal bone respectively, while ($b_1$), ($b_2$), and ($b_3$) have the regions of facial nerve marked in red. The two enlarged images in Fig. 3($a_2$') and ($b_2$') display the enlargement of the axial view and the corresponding ground truth. Due to the small size of the facial nerve and the limitation of the GPU memory, a $64 \times 64 \times 32$ block containing a part of the facial nerve is cropped and used as the input for the network. To increase the training dataset and alleviate overfitting, the training data are transformed by the concatenation of random cropping, scaling, rotating, and mirroring.

### B. Implementation Details and Evaluation Metrics

Our UADSN model is implemented using PyTorch on an NVIDIA-RTX-4090 GPU. The AdamW optimizer is utilized with learning rate, betas, and weight decay set to 1e-3, $(0.9, 0.999)$, 5e-4, respectively. We configure the training epochs as 50k, batch size as 4, and loss weight $\alpha$ as 0.5.

We evaluate the proposed UADSN using three metrics: Dice similarity coefficient (DSC), Average Symmetric Surface Distance (ASSD), and Average Hausdorff Distance (AHD). DSC is a measure of set similarity, which is more sensitive to the internal filling of segmentation. The DSC value ranges from 0 to 1, where the higher value indicates the better segmentation performance. We denote the prediction as $P$ and the ground truth as $G$, and the DSC can be computed as follows:

$$DSC(P, G) = \frac{2|P \cap G|}{|P| + |G|}, \quad (9)$$

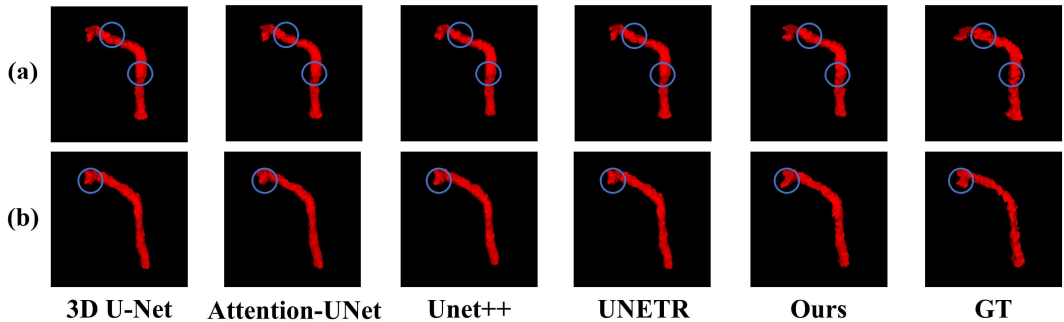

Fig. 4. Visualized segmentation results of various approaches on our facial nerve dataset. The rows (a) and (b) correspond to different samples.

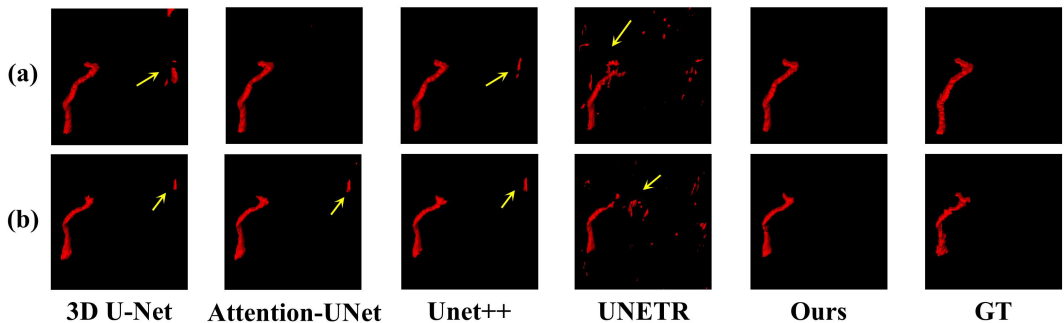

Fig. 5. Visualized results of different approaches without using the maximum region growing method.

where $|\cdot|$ denotes the number of foreground voxels.

ASSD and AHD are surface-based metrics that are more sensitive to the prediction boundary. Low ASSD and AHD indicate better segmentation performance. We use $S(P)$ to represent the set of surface voxels of $P$. The shortest distance of an arbitrary voxel $v$ to $P$ is defined as:

$$d(v, S(P)) = \min_{s_P \in S(P)} \|v - s_P\|, \quad (10)$$

where $\|\cdot\|$ represents the Euclidean distance. Based on this, the ASSD can be defined as follows:

$$ASSD(P, G) = \frac{1}{|S(P)| + |S(G)|} \times \left[ \sum_{s_P \in S(P)} d(s_P, S(G)) + \sum_{s_G \in S(G)} d(s_G, S(P)) \right] \quad (11)$$

The unidirectional Average Hausdorff Distance from $P$ to $G$ is defined as:

$$d(P, G) = \frac{1}{N} \sum_{p \in P} \min_{g \in G} \|p - g\|, \quad (12)$$

where $N$ is the number of voxels in $P$. Then AHD can be formulated as:

$$AHD(P, G) = \max(d(P, G), d(G, P)). \quad (13)$$

TABLE I. COMPARISONS WITH OTHER APPROACHES ON OUR FACIAL NERVE DATASET.

| Method | Metrics | | |
|---|---|---|---|
| | DSC (%) | ASSD (mm) | AHD (mm) |
| 3D U-Net [7] | 77.48 | 0.1250 | 0.1638 |
| Attention U-Net [8] | <u>78.43</u> | <u>0.1130</u> | <u>0.1458</u> |
| UNet++ [9] | 77.82 | 0.1223 | 0.1778 |
| UNETR [10] | 63.16 | 0.4829 | 0.7635 |
| UADSN (Ours) | **79.79** | **0.0952** | **0.1264** |

### C. Comparative Experiments Results

We compare the proposed UADSN with four famous segmentation approaches (i.e., 3D U-Net [7], Attention U-Net [8], UNet++ [9], and UNETR [10]) on our self-built facial nerve dataset. Table I presents the quantitative results of various approaches. The optimal results are in bold, and the second-ranked results are underlined. Our UADSN obtains the highest DSC of 79.79% and the lowest ASSD and AHD, exhibiting a 1.36% DSC improvement compared to the second-ranked method.

Fig. 4 shows some visualized facial nerve segmentation results. To eliminate misclassified regions, we use the maximum region growing method as the post-process for all the approaches. It can be seen that our UADSN can achieve a more complete structure, highlighted by blue circles. One of the main reasons is that our uncertainty-aware consistency loss makes the network focus on regions that are difficult to segment.

TABLE II. ABLATION STUDIES ON OUR FACIAL NERVE DATASET.

| Method | | | Metrics | | |
|---|---|---|---|---|---|
| clDice Loss | sSE | Dual-Stream | DSC (%) | ASSD (mm) | AHD (mm) |
| | | | 77.64 | 0.1268 | 0.1743 |
| √ | | | 79.01 | <u>0.0993</u> | 0.1360 |
| √ | √ | | <u>79.07</u> | 0.1022 | <u>0.1328</u> |
| √ | √ | √ | **79.79** | **0.0952** | **0.1264** |

To examine whether the segmentation method tends to misclassify non-facial nerve regions as facial nerves, we compare the visualized results of different methods without using the maximum region growing method in Fig. 5. It is easily seen that there are many false positive predictions in the segmentation results of other methods, highlighted by the yellow arrows. Even without post-processing, our UADSN hardly misidentifies non-facial nerve regions as facial nerves. The key factor is that we employ the sSE module in both the 2D and 3D segmentation streams, emphasizing important spatial information. Additionally, the use of clDice loss improves the topology precision of the predictions and suppresses misidentification.

*D. Ablation Studies*

In order to demonstrate the effect of each component, we conduct ablation studies on our facial nerve dataset. As shown in Table II, we only use Dice loss to train a 3D U-shaped CNN and obtain a DSC of 77.64%. When we introduce clDice loss into the supervised loss, the DSC increases by 1.37%, the ASSD decreases by 0.0275mm, and the AHD decreases by 0.0383mm. This demonstrates the importance of topology-preservation for tubular organ segmentation. The addition of sSE modules reduces false positive predictions, obtaining a 0.06% improvement in DSC. When we introduce the 2D segmentation stream and uncertainty-aware consistency loss into the network, we achieve optimal performance, indicating the complementarity of our submodules. The consistency loss between the two segmentation streams improves accuracy in difficult-to-segment regions.

V. CONCLUSION

In this paper, we propose UADSN, an uncertainty-aware dual-stream network for facial nerve segmentation in CT scans. UADSN contains two segmentation streams (i.e., the 3D segmentation stream and the 2D segmentation stream), and an uncertainty-aware consistency loss is employed to achieve fine-grained segmentation. Besides, sSE modules are introduced into the network to emphasize significant spatial information. Furthermore, we use a clDice loss in the supervised loss to explore the importance of topology-preservation for facial nerve segmentation. Experiments on the facial nerve dataset indicate the superior performance of UADSN compared with other famous approaches. We hope that UADSN can inspire further research in the segmentation of tubular organs. In the future, we will explore novel methods to calculate uncertainty and further enhance the performance of UADSN.

ACKNOWLEDGMENT

The authors would like to express thanks to Professor Yu-Tang Ye and the staff at the MOEMIL laboratory for their guidance on this project. At the same time, they thank the First Affiliated Hospital of Guangzhou Medical University for providing temporal bone CT images with annotations. This work was funded in part by the Fundamental Research Funds for the Central Universities of China from the University of Electronic Science and Technology of China under Grant ZYGX2019J053 and Grant ZYGX2021YGCX020.